\documentstyle[11pt]{article}
\newcommand{\papertitle}[1]{{\Large {\bf #1}}\vspace{3mm}}
\newcommand{\paperauthor}[1]{\large {\bf #1}\\\vspace{2mm}}
\newcommand{\address}[1]{{\it #1}}
\def\abstractname{Abstract.\hspace{2mm}}
\def\abstract{%
\list{}{\advance\topsep by 0.35cm
\footnotesize
 \leftmargin=1.5cm
 \labelsep=0mm
\labelwidth=0mm
 \rightmargin\leftmargin}\item[\bf\abstractname]}

\newcommand{\reduction}[2]{\left.\phantom{\bigl|} #1 \right|_{#2}}
\newcommand{\Real}{\mathop{\rm Re}}

\newcommand{\C}{{\if mm {{\rm C}\mkern -15mu{\phantom{\rm t}\vrule}}
\mkern +10mu \else \leavemode \hbox{I}\kern -.17em \hbox{C} \fi}}
\begin{document}
\begin{center}
\papertitle{THREE--BODY RESONANCES
IN FRAMEWORK OF THE FADDEEV
CONFIGURATION SPACE APPROACH%
\footnote{LANL E-print {\tt nucl-th/9702037}. To appear in 
Proc. of Intern. Conf. on Computational Modelling 
and Computing in Physics (Dubna, September 16--21, 1996).}}
\\
\paperauthor{
E. A. Kolganova, A. K. Motovilov
}
\address{Joint Institute for Nuclear Research, Dubna, Russia \\
e-mail: kea@thsun1.jinr.ru, motovilv@thsun1.jinr.ru}
\end{center}

\begin{abstract}
Algorithm, based on explicit representations for analytic 
continuation of the T-matrix Faddeev components on unphysical 
sheets, is worked out for calculations of resonances in the 
three-body quantum problem. According to the representations, 
poles of the T--matrix, scattering matrix and resolvent on 
unphysical sheets, interpreted as resonances, coincide with 
those complex energy values where appropriate truncations of the 
scattering matrix have zero as eigenvalue. Scattering amplitudes 
on the physical sheet, necessary to construct scattering matrix, 
are calculated on the basis of the Faddeev differential 
equations.  Effectiveness of the algorithm developed is 
demonstrated for example of searching for resonances in the 
system $nnp$ and in a model three-boson system.
\end{abstract}

We make a numerical test of the approach proposed in 
Ref.~\cite{1} to treat the three--body resonances in the case of 
pairwise interactions falling off in coordinate space not slower 
than exponentially.  This approach is based on a construction of 
explicit representations for the Faddeev components 
$M_{\alpha\beta}(z)$, $\alpha,\beta=1,2,3,$ of the three--body 
T-matrix ${T}(z)$ as well as for the scattering matrices 
${S}(z)$ and resolvent ${R}(z)$ continued on unphysical sheets 
of the energy $z$ plane.  For the sheets, the notation $\Pi_l$ 
is used with $l$, an enumerating multi--index (see~\cite{1}). 
The representations constructed demonstrate a structure of 
kernels of the above operators after continuation and give new 
capacities for analytical and numerical studies of three--body 
resonances. In particular the representations for analytic 
continuation $\reduction{M(z)}{\Pi_l}$ of the matrix 
$M(z)=\{M_{\alpha\beta}(z)\}$ on the sheet $\Pi_l$ with details 
omitted read 
$$
\reduction{M}{\Pi_l}=\reduction{M}{\Pi_0}-
Q^{\dagger}_l{\bf J}^\dagger AS^{-1}_l{\bf J}Q_l
$$
The operator $Q_l(z)$ and the ``transposed'' one, 
$Q_l^\dagger(z)$ are obviously constructed of the matrix $M(z)$ 
taken on the physical sheet $\Pi_0$.  $A(z)$ is a number matrix, 
an entire function of $z\in{\C}$.  By $S_l(z)$ we understand 
a truncation (depending essentially on $l$) of the total 
three--body scattering matrix $S(z)$. Operators ${\bf J}(z)$ and 
${\bf J}^\dagger(z)$ realize a restriction of kernels of the 
operators $Q_l(z)$ and $Q_l^\dagger(z)$ on the energy shells, 
respectively, in first and last momentum variables.  So that the 
products $Q_l^\dagger{\bf J}^\dagger$ and ${\bf J}Q_l$ have 
half--on--shell kernels.  Representations~\cite{1} for 
analytical continuation of the three--body scattering matrices 
and resolvent follow immediately from the representations above 
for $\reduction{M(z)}{\Pi_l}$.

As follows from the representations constructed, the nontrivial (i.e.
differing from the poles at the discrete spectrum eigenvalues of the
three--body Hamiltonian) singularities of the T--matrix, scattering
matrices and resolvent situated on an unphysical sheet $\Pi_l$, are
singularities of the inverse truncated scattering matrix
$S_l^{-1}(z)$.  Therefore, the resonances on the sheet $\Pi_l$
considered as poles of the T--matrix, scattering matrix and
resolvent continued on $\Pi_l$  are those values of the energy
$z$ for which the matrix $S_l(z)$ has zero as eigenvalue.
Thereby, to search for the resonances situated on a certain 
unphysical sheet $\Pi_l$, one can apply any method allowing to 
compute analytical continuation on the physical sheet of the 
elastic scattering, rearrangement or breakup amplitudes 
necessary for construction of respective ${S}_l(z)$.  In 
particular such is the algorithm developed for $(2\rightarrow 
2,3)$ processes in Ref.~\cite{2} on the base of Faddeev 
differential formulation of the scattering problem in 
configuration space (see also~\cite{3,4} and Refs. therein).  It 
is only necessary to go out in this formulation on the complex 
plane of $z$ including the asymptotical boundary conditions.

In the present work we utilize a code based on the 
algorithm~\cite{2}--\cite{4}, for computations of s-state  $nnp$ 
resonances situated on the unphysical sheet $\Pi_{(0,1)}$ 
connected with physical one by crossing the spectral interval 
$(E_d,\, 0)$ between the deuteron energy $z\!  =\! E_d$ and 
breakup threshold $z\! =\! 0$.  We solve the two--dimensional 
Faddeev integro--differential equations~\cite{4} with the 
$(2\rightarrow 2,3)$ asymptotical boundary 
conditions~\cite{2}--\cite{4} at complex energies $z$ and 
extract the truncated s-state scattering matrix $S_{01}(z)\! =\! 
1 +2ia_0(z)$ with $a_0(z)$, the amplitude of elastic $nd$ 
scattering continued on the physical sheet.  When making a 
finite--difference approximation of the equations above in polar 
coordinates we take up to 180 points of grids in both radial and 
angular variables, the cut--off radius being up to 39~fm.  As a 
{\it NN}--interaction, the Malfliet--Tjon potential 
MT~I--III~\cite{5} is chosen.
\begin{figure}
\centering
\unitlength=0.24pt         
%
\begin{picture}(999,1100)
\put(-300,1100){\special{em:graph sm1ac.pcx}}
\end{picture}
\caption{ Surface of the function $|S_{01}(z)|$ in the model
system of three bosons with the nucleon masses.
The potential
$V^G(r)$ is used with the barrier $V_b=1.5$~MeV.
Position of the resonance  $z_{\rm res}(\mbox{3B})$
corresponds to the minimal (zero) value of
$|S_{01}(z)|$.
}
\label{fig-surface}
\end{figure}
\begin{figure}
\centering
\unitlength=0.24pt         
%
\begin{picture}(999,700)
\put(-0,700){\special{em:graph tr5.pcx}}
\end{picture}
\caption{
Trajectory of the resonance $z_{\rm res}(\mbox{3B})$ on the 
sheet $\Pi_{(0,1)}$ in the model system of three bosons with the 
nucleon masses. The potential $V^G(r)$ is used. Values of the 
barrier $V_b$ in MeV are given near the points marked on the 
curve.
}
\label{fig-trajectory1}
\end{figure}

Firstly, we have checked a validity of the code finding the $^3\! H$
bound--state energy $E_t$ as a pole of the function $S_{01}(z)$. More
precisely, the location of the $S_{01}(z)$ pole was found as a root
of the inverse amplitude $1/a_0(z)$.  Beginning from the grid dimension
$80\!\times\! 80$, we have obtained  $E_t\! =\! -8.55$~MeV.
Hereafter all the energies are given with respect to the breakup
threshold.  Note that the value stated is in a good agreement with
known results on $E_t$ in the MT~I--III model (see Ref.~\cite{6}).

Concerning the $nnp$ resonances on the sheet $\Pi_{(0,1)}$, we 
have inspected a domain of a range about $10$~MeV in vicinity of 
segment $[E_d,0]$ in the complex $z$ plane. Especially carefully 
we studied a vicinity of the points $z\! =\! -1.5\pm 
0.3+i(0.6\pm 0.3)$~MeV interpreted in Refs.~\cite{7,8} as a 
location of an exited state energy of \, $^3\! H$.  
Unfortunately we have succeeded to find only one root $z_{\rm 
res}$ of the function $S_{01}(z)$, corresponding to the known 
virtual state of the $nnp$ system at total spin S=1/2.  On a 
$180\!\times\! 180$ grid, we have found \underline{$z_{\rm 
res}\! =\! -2.728$~MeV} i.e. it is situated $0.504$~MeV to the 
left from the $nd$ threshold $E_d\! =\! -2.224$~MeV (in the 
MT~I--III model).  Note that the shift $E_d\! -\! z_{\rm res}$ 
found from experimental data on $nd$ scattering, is $0.515$~MeV 
(see Ref.~\cite{9}).  Its value~\cite{9} computed in a 
separabilized MT~I--III model on the base of the momentum space 
Faddeev equations, is equal to $0.502$~MeV. As one could expect 
(see the data on three-nucleon resonances in~\cite{9}), we have 
failed to find any resonances in the quartet state at $L=0$ as 
well as at $L=1$.

Also, we have studied a behavior of a resonance situated on the
unphysical sheet $\Pi_{(0,1)}$ in a model three--body system including
bosons with masses of the nucleon.
As a pairwise interaction
between the bosons we have used the Gauss-type potential
supplied with an additional Gauss repulsive barrier term,
$$
    V^G(r)=V_0 \exp[-\mu_0 r^2] + V_b \exp[-\mu_b (r-r_b)^2]
$$
where the values $V_0=-55$~MeV, $\mu_0=0.2$~fm$^{-2}$,
$\mu_b=0.01$~fm$^{-2}$,  $r_b=5$~fm have been fixed
while the barrier amplitude
$V_b$ varied.  A resonance (with non-zero imaginary part) on the
sheet $\Pi_{(0,1)}$ arises in the system concerned just due to
the presence of the barrier term.  Example of a surface of the
$|S_{01}(z)|$ function for the barrier amplitude
$V_b=1.5$~MeV is shown in Fig.~\ref{fig-surface} (for a
$80\!\times\! 80$~grid).  A trajectory of the resonance $z_{\rm
res}(\mbox{3B})$ (a zero of the function $S_{01}(z)$) is
shown for the changing barrier $V_b$ in
Fig.~\ref{fig-trajectory1}.  This trajectory was watched for the
barrier $V_b$ decreasing in the interval between $1.5$~MeV and
$0.85$~MeV.
\begin{figure}
\centering
\unitlength=0.24pt         
%
\begin{picture}(800,700)
\put(-100,700){\special{em:graph td5.pcx}}
\end{picture}
\caption{
Dependence of the ``deuteron'' energy $E_d$ (curve~1)
and real part of the resonance
$z_{\rm res}(\mbox{3B})$ (curve~2) on the barrier value $V_b$.
}
\label{fig-trajectory2}
\end{figure}
When drawing the trajectory,  we have used a $160\!\times\!  
160$~grid.  It can be seen from Fig.~\ref{fig-trajectory1} that 
the behavior of the resonance $z_{\rm res}(\mbox{3B})$ is rather 
expected:  with monotonously decreasing real part, the imaginary 
part of the resonance changes also monotonously.  For $V_b \geq 
0.95$~MeV the energy $z_{\rm res}(\mbox{3B})$ becomes real, 
turning into a discrete spectrum eigenvalue.  In 
Fig.~\ref{fig-trajectory2} we plot both the trajectories of the 
resonance real part $\Real z_{\rm res}(\mbox{3B})$ and two-boson 
binding energy $E_d$.  As one can see from this figure, the 
value of $|\Real z_{\rm res}(\mbox{3B})|$ increases more quickly 
than $|E_d|$, coinciding with $|E_d|$ at $V_b \cong 0.97$~MeV.  

Trajectory of the resonance concerned in the lower complex 
half-plane is symmetric to the curve shown in 
Fig.~\ref{fig-trajectory1} with respect to the real axis. 
Respective points, symmetric to those marked in 
Fig.~\ref{fig-trajectory1}, correspond to the same values of 
$V_b$.

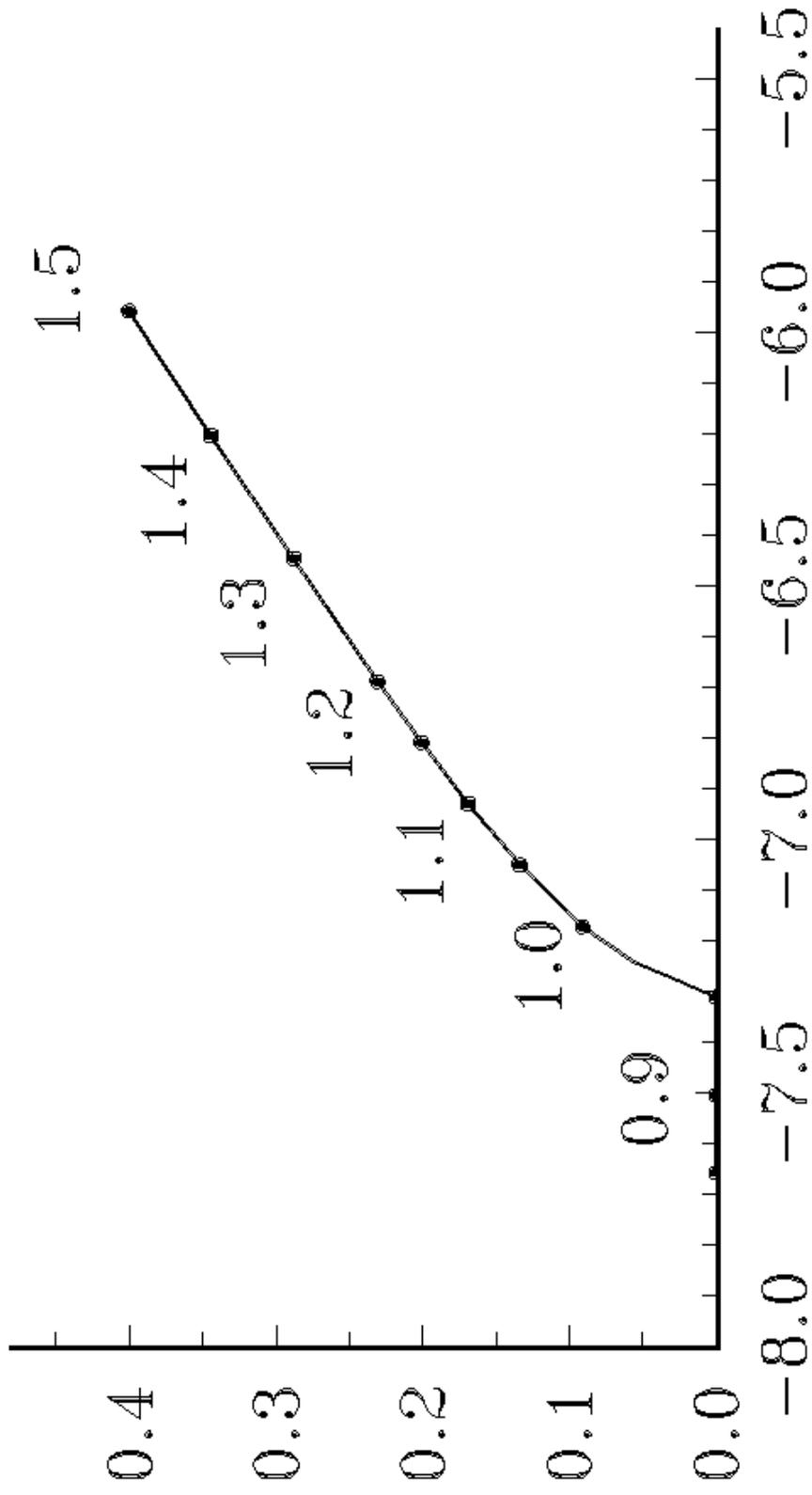
\end{document}